\documentclass[twocolumn]{aastex701}

\usepackage{ulem}
\usepackage{bm}
\usepackage{amsmath}
\usepackage{enumitem}

\renewcommand{\sc}{{\rm sc}}

\begin{document}

\title{Negative and Positive Cascade Rates of Slow Alfv\'enic Turbulence in Switchback and Non-switchback Intervals: \emph{Wind} Observations}


 \correspondingauthor{Sofiane Bourouaine}
\email{sbourouaine@fit.edu}

\author[0000-0002-2358-6628]{Sofiane Bourouaine}
\email{sbourouaine@fit.edu}
\affiliation{%
 Department of Aerospace, Physics and Space Sciences, Florida Institute of Technology, Melbourne, Florida, 32901, USA.
}%

\begin{abstract}
We estimate the turbulent cascade rate in imbalanced Alfv\'enic turbulence in the slow solar wind using a Politano--Pouquet-based approach that does not require an isotropic assumption. We analyze selected 8-hour intervals provided by \emph{Wind} spacecraft. These intervals are characterized by near-homogeneity, weak velocity shear, and imbalanced Alfv\'enic fluctuations, and are classified as switchback or non-switchback depending on the angular spread of the magnetic field relative to the mean field.
Switchback intervals are identified by large magnetic-field rotations, whereas non-switchback intervals exhibit only small angular deviations. Our results show that non-switchback intervals exhibit a positive total cascade rate and power spectra consistent with a $-3/2$ scaling, indicative of a forward energy cascade. In contrast, switchback intervals are characterized by a negative total cascade rate (inverse cascade), and by steeper power spectra with a spectral index of approximately $-1.6$.
This indicates that large magnetic-field rotations can substantially modify the scale-to-scale energy transfer in Alfv\'enic slow solar wind. The results suggest that switchbacks may be associated with additional physical processes, such as local instabilities, or reconnection-related structures, that alter the direction and magnitude of the turbulent cascade. These findings may help explain previous reports of negative cascade rates in imbalanced Alfv\'enic solar-wind turbulence.
\end{abstract}



\section{Introduction}

In-situ measurements have shown that the solar wind is a turbulent plasma
\citep{verscharen19}. Fast solar wind, typically characterized by speeds
$U\gtrsim500$ km~s$^{-1}$, is generally weakly compressible and dominated by imbalanced
Alfv\'enic fluctuations that propagate mainly outward from the Sun
\citep{bruno05}. In contrast, slow solar wind, with speeds
$U\lesssim500$ km~s$^{-1}$, exhibits a broader range of turbulent properties. Some slow-wind
intervals are more compressible and less Alfv\'enic, whereas others display properties
similar to those of fast solar wind \citep[see e.g.,][]{damicis15,stansby20,chen20,dorseth24b}.
The dissipation of turbulent fluctuations is thought to contribute to the heating of the
solar-wind plasma \citep{vasquez07,stawarz09,smith09,perez13,perez21b,bourouaine24}.

In the standard picture of magnetohydrodynamic (MHD) turbulence, energy injected at
large scales is transferred through the inertial range toward progressively smaller scales.
At kinetic scales, part of this turbulent energy can be dissipated and converted into ion
and electron heating. Therefore, a forward turbulent cascade is an essential ingredient
for understanding the thermal evolution and heating of the solar wind.
In the incompressible MHD turbulence, the scale-to-scale
transfer of energy can be quantified using exact third-order relations. In particular,
\cite{PP98} (hereafter PP98) derived an exact law for homogeneous incompressible MHD
turbulence, which relates mixed third-order structure functions of the Els\"asser variables
to the mean turbulent energy cascade rate.

Several observational studies have used the PP98 relation to estimate the
cascade rate in the solar wind. Using single-spacecraft measurements from the
\emph{Advanced Composition Explorer} (ACE), \citet{macbride05} estimated the cascade
rate by assuming isotropic symmetry for the third-order structure functions. Later
studies introduced axisymmetric assumptions, which are more appropriate for magnetized solar-wind turbulence~\citep{boldyrev09}, and applied this approach to different solar-wind conditions \citep{macbride08,stawarz09,stawarz10,stawarz11}. The turbulent cascade rate has also been estimated using multi-spacecraft measurements in the solar wind
and in near-Earth plasma environments \citep{osman11,bandyopadhyay18}. In addition,
extensions of the exact-law formalism have been developed to account for compressible
effects and have been applied to compressible solar-wind turbulence
\citep{banerjee13,andres17,hadid17}.

Although the classical turbulent-cascade picture predicts a net forward transfer of
energy, several studies have reported negative values of the total or
average cascade rate in some solar-wind intervals
\citep{smith09,stawarz09,stawarz10,vasquez18}. A negative cascade rate is usually
interpreted as evidence for a back-transfer, or inverse cascade, of turbulent energy from
smaller scales to larger scales. Such behavior may indicate that the standard
forward turbulent-heating pathway is locally reduced, modified, or influenced by
additional physical processes. Negative cascade rates are not generally expected in the
standard phenomenology of homogeneous incompressible MHD turbulence, nor are they
observed in conventional numerical simulations of MHD turbulence
\citep{muller05,perez12}. Observationally, negative cascade rates have been
associated with intervals of high cross-helicity \citep{smith09,stawarz10}, as well as
with intervals characterized by decreasing solar-wind speed \citep{stawarz11,vasquez18}.

In this work, we investigate the turbulent cascade rate in imbalanced Alfv\'enic slow solar wind turbulence, and we focus on how the high prevalence of large field rotations (or switchbacks) affects the cascade rate. Here, we use the term ``Alfv\'enic slow solar wind'' to refer specifically to slow-wind intervals characterized by fluctuations showing relatively constant magnetic-field magnitude, and relatively weak density fluctuations \citep[see, e.g., ][]{dmicis21, dorseth24b}.

 The presence of switchbacks in the solar wind has been known for decades~\citep{balogh99,kahler96}, but they have been more extensively investigated recently using \emph{Parker Solar Probe} measurements \citep{bale19,kasper19,td20,horbury20,mozer20}.
These structures are commonly interpreted as sharp Alfv\'enic impulses or magnetic-field
folds embedded in the solar wind \citep{matteini14,tenerani20,horbury20,bourouaine20,bourouaine22}. Several mechanisms have been proposed for their formation, including nonlinear evolution of large-amplitude
Alfv\'en waves, turbulent generation, magnetic-field folding, interchange reconnection,
and shear-driven or expansion-related processes
\citep{squire20,mallet21,shoda21,drake21,zank20,schwadron21,Ruffolo20}.

Understanding the role of switchbacks in imbalanced Alfv\'enic turbulence is important
because large magnetic-field rotations may affect both the magnitude and the direction
of the turbulent energy transfer. If switchbacks are associated with additional
nonlinear structures, instabilities, or reconnection-related processes, then intervals
containing switchbacks may not follow the same cascade behavior as more standard
Alfv\'enic intervals without large field rotations. This distinction may help explain why
some previous studies found negative cascade rates in highly Alfv\'enic solar-wind
intervals. In the present study, we therefore separate selected slow-wind Alfv\'enic
intervals into switchback and non-switchback populations and estimate the cascade
rate using a reduced Politano--Pouquet-based approach that does not require an
isotropic assumption.

In the next section, we describe the theoretical background and data-analysis methodology. In Section~3, we present the main results, and in Section~4, we discuss and summarize our findings.

\section{Methodology and data analysis \label{sec:methodology}}
In this study, we use plasma and magnetic-field measurements from the \emph{Wind} spacecraft  \citep{lepping95} over a period of approximately three years, from January 1, 2017, to December 31, 2019.The data were obtained through the NASA Coordinated Data Analysis Web (CDAWeb). The plasma and magnetic-field data are originally available at a temporal resolution of 3~s \citep{Lin2021Wind3DP}. To ensure a common time base and improve the synchronization among the proton number density, bulk velocity, and magnetic-field vectors, all quantities are resampled to a uniform cadence of 10~s. All vector quantities are expressed in the Geocentric Solar Ecliptic (GSE) coordinate system. In this system, the $x$-axis points from Earth toward the Sun, the $y$-axis lies in the ecliptic plane and points toward dusk, opposite to the direction of planetary motion, and the $z$-axis points toward ecliptic north.

\subsection{Selection of Intervals}
The dataset is divided into 8-hour intervals, and only slow-wind intervals (mean bulk speed $V_0\lesssim500$ km~s$^{-1}$) are considered. We then apply a set of selection criteria to retain intervals that are approximately homogeneous, weakly compressible, have low velocity shear, and have imbalanced turbulence.

First, for each interval we compute the mean magnetic field, $\mathbf{B}_0$, and the most probable magnetic field, $\mathbf{B}_m$ (obtained from the most probable values of the vector components $(B_x,B_y,B_z)$ of the instantaneous magnetic field ${\bf B}$). To avoid intervals containing mixed magnetic polarities, such as those associated with heliospheric current-sheet crossings, we require that $\mathbf{B}_0$ and $\mathbf{B}_m$ are nearly identical. This condition is imposed if $|\phi_0-\phi_m|<10^\circ$, where $\phi_0$ ($\phi_m$) is the angle between the x-axis and $\mathbf{B}_0$ ($\mathbf{B}_m$). We also require the angular difference to satisfy $|\psi_0-\psi_m| < 10^\circ$, where $\psi_0$ and $\psi_m$ denote the angles between the $z$-axis and ${\bf B}_0$ and ${\bf B}_m$, respectively.

Second, we retain only intervals with low values of plasma compressibility, $C_n$, and magnetic compressibility, $C_B$, defined as
\begin{equation}
C_n=\frac{\langle (n-n_0)^2\rangle}{n_0^2},
\qquad
C_B=\frac{\langle (B-B_0)^2\rangle}{B_0^2},
\end{equation}
where $n_0$ and $B_0$ are the mean proton number density and mean magnetic-field magnitude, respectively, while $n$ and $B$ are their instantaneous values. We require that $C_n$ and $C_B$ be less than 0.15.

Third, we select intervals with high normalized cross-helicity, $|\sigma_c|>0.5$, where
\begin{equation}
\sigma_c=
\frac{E^{\mathrm{+}}-E^{\mathrm{-}}}
{E^{\mathrm{+}}+E^{\mathrm{-}}},
\end{equation}
where $E^{\pm}=\left\langle |\delta\mathbf{z}^{\pm}|^2 \right\rangle$, and the Els\"asser fluctuations are defined as $\delta\mathbf{z}^{\pm}=\delta\mathbf{V}\pm\delta\mathbf{b}$ with $\delta\mathbf{V}=\mathbf{V}-\mathbf{V}_0$ is the fluctuating bulk velocity, and
$\delta\mathbf{b}=\mathbf{b}-\mathbf{b}_0$ is the fluctuating magnetic field. Here, $\mathbf{V}_0$ and $\mathbf{b}_0$ are interval averages, and
$\mathbf{b}=\mathbf{B}/\sqrt{\mu_0 m_p n_0}$ is the magnetic field expressed in Alfv\'en-speed units, with $\mu_0$ the vacuum permeability and $m_p$ the proton mass.

Finally, we require the selected intervals to have homogeneous solar wind speed (or weak velocity shear). This condition is quantified using
\begin{equation}
C_V=\frac{\langle (V-V_0)^2\rangle}{V_0^2},
\end{equation}
where $V=|\mathbf{V}|$ and $V_0$ is its interval average. Only intervals satisfying $C_V<0.15$ are retained.

Together, these criteria ensure that the selected intervals are suitable for applying the Politano--Pouquet exact law to estimate the turbulent energy cascade rate. Over the 3-year period analyzed, we identified 75 8-hour intervals that satisfy all selection criteria described above.

\subsection{Selection of switchback and non-switchback intervals}
Among the selected Alfv\'enic intervals, we further classify them according to the angle associated with magnetic-field rotation. In particular, we distinguish intervals containing large magnetic-field rotations from those characterized by relatively small rotations. Previous studies have shown that switchback intervals are associated with magnetic-field folding, kinks, or reversals, which can produce large angular deviations of the instantaneous magnetic field from the background field \citep{td20,bourouaine20,bourouaine22,jagarlamudi23}. In such intervals, the angle $\phi_B$ between the instantaneous magnetic-field vector, $\mathbf{B}$, and the background magnetic field, $\mathbf{B}_0$, can exceed $90^\circ$. For example, close to the Sun, where $\mathbf{B}_0$ is nearly radial, a switchback can cause the radial component of $\mathbf{B}$ to reverse sign. This behavior is not expected in non-switchback intervals, where the magnetic field remains relatively well aligned with the background field.

To quantify the angle of the magnetic-field rotation around the mean field, we compute the normalized histogram $H(\phi)$ of the magnetic-field angle $\phi$, where $\phi$ is the angle between the GSE $x$-axis and the instantaneous magnetic-field vector $\mathbf{B}$. From $H(\phi)$, we identify two angles, $\phi_a<\phi_0$ and $\phi_b>\phi_0$, at which the normalized histogram decreases to 0.1. Here, $\phi_0$ is the angle associated with the mean magnetic field $\mathbf{B}_0$. The angles $\phi_a$ and $\phi_b$ therefore define the lower and upper bounds of the dominant angular distribution around $\phi_0$. Magnetic-field rotations outside this range occur with a normalized probability below 0.1 and are therefore considered statistically insignificant for the classification.

Numerical simulations of homogeneous reduced MHD turbulence suggest that substantial magnetic-field rotations relative to the local mean magnetic field, $\mathbf{B}_0$, are generally limited to angular deviations of order $\phi \lesssim 25^\circ$. This estimate follows from the fluctuation amplitudes commonly adopted in reduced-MHD simulations, where the normalized parameters in the simulation are the rms magnetic fluctuation is $B_{\rm rms}\sim 1$ and the guide field to $B_0=5$ \citep{perez12}. These values correspond to an rms angular deviation of
$\phi_{\rm rms} \simeq \tan^{-1}\!\left(B_{\rm rms}/B_0\right)
\simeq 11^\circ$. Assuming a Gaussian distribution of magnetic-field rotation angles, the probability density decreases to approximately 10\% of its peak value at $\phi_{\rm rms} \simeq 25^\circ$. Thus, magnetic-field rotations significantly exceeding $\sim 25^\circ$ are expected to be relatively uncommon in homogeneous reduced-MHD turbulence.

Guided by this result, intervals satisfying $|\phi_a-\phi_0|<30^\circ$ and $|\phi_b-\phi_0|<30^\circ$ are classified as non-switchback intervals, representing cases with relatively small magnetic-field rotations. Conversely, intervals satisfying $|\phi_a-\phi_0|>40^\circ$ or $|\phi_b-\phi_0|>40^\circ$ are classified as switchback intervals, representing cases with large magnetic-field rotations.

\begin{figure*}[!t]
    \centering
    \includegraphics[]{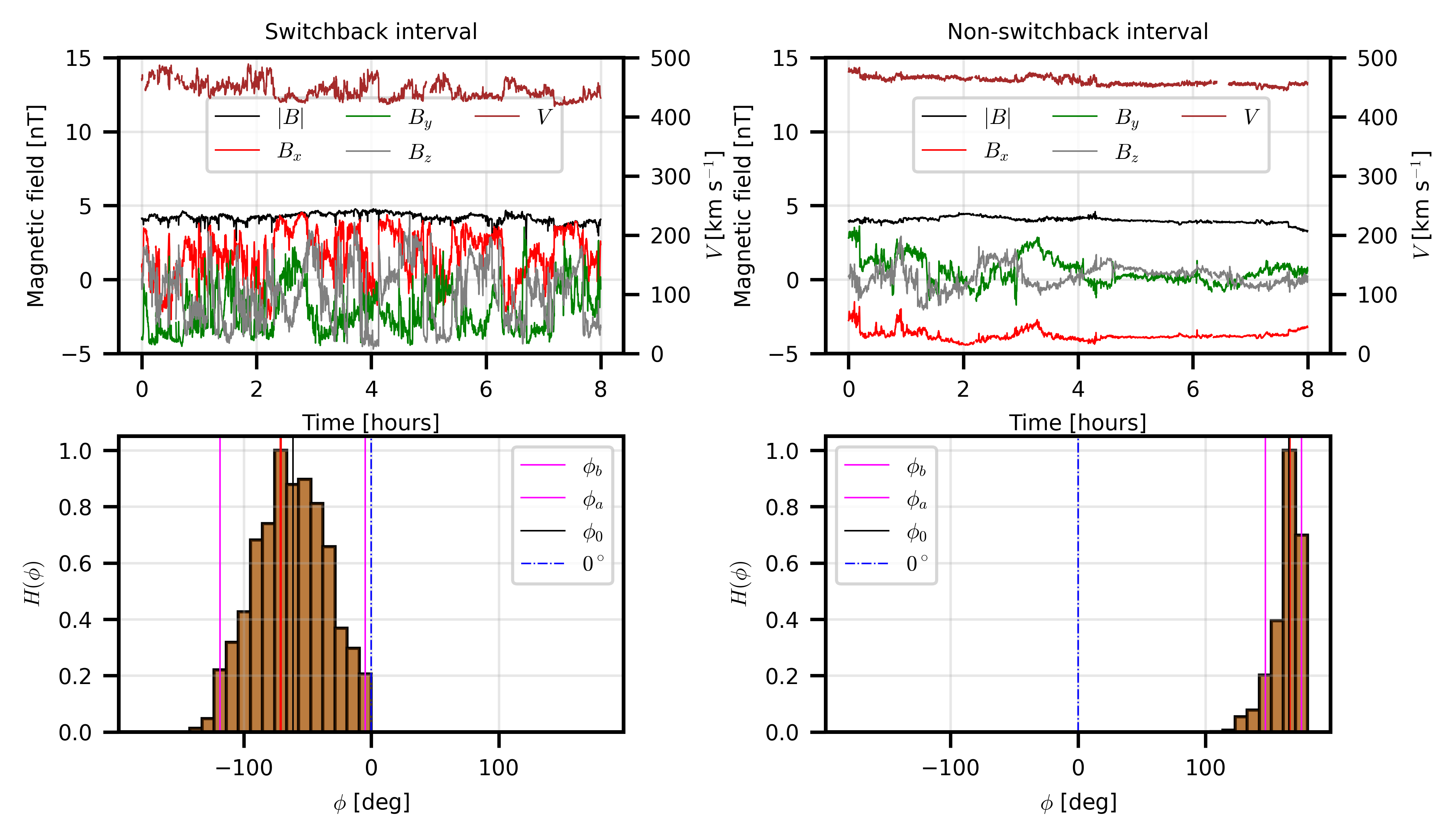}
     
     \vspace*{-0.1cm}
   \caption{Top panels show the magnetic-field components, magnetic-field magnitude  (left $y-$axis), and bulk solar-wind speed (right $y-$axis) for a representative switchback interval (left) and non-switchback interval (right). Bottom panels show the normalized histogram of the magnetic-field angle $\phi$ for the same intervals. The black vertical line marks $\phi_0$, the angle of the mean magnetic field $\mathbf{B}_0$, and the red vertical line marks the angle of the most probable magnetic-field direction, $\mathbf{B}_m$. The magenta vertical lines indicate $\phi_a$ and $\phi_b$, which define the angular extent of the dominant distribution around $\phi_0$.}
    \label{fig:fig1}
\end{figure*} 

Figure~\ref{fig:fig1} shows an example of the magnetic-field rotation analysis for a switchback interval (left panels) and a non-switchback interval (right panels). Both intervals satisfy the selection criteria described above, including approximate homogeneity, weak compressibility, low velocity shear, and high Alfv\'enicity. The bulk speed $V$, proton number density, $n$, and magnetic-field magnitude, $B$ remain relatively uniform throughout both intervals.  The mean plasma and turbulence parameters for the switchback (non-switchback) interval are $V_0 = 480$ ($460$)~km~s$^{-1}$, $n_0 = 3.2$ ($2.5$)~cm$^{-3}$, $\langle B\rangle = 4.2$ ($4.0$)~nT, $C_n = 0.07$ ($0.10$), $C_B = 0.07$ ($0.05$), and $C_V = 0.03$ ($0.02$), respectively.


In both examples, the angle of the mean magnetic field, $\phi_0$, is close to the angle of the most probable magnetic-field direction, $\phi_m$, with a difference less than $10^\circ$. This indicates that both intervals have a well-defined background magnetic-field direction.

The histogram of $\phi$ for the switchback interval exhibits a broad angular distribution, indicating large rotations of the instantaneous magnetic field around the mean field. In this case, the angular bounds satisfy $|\phi_0-\phi_a|\simeq 60^\circ$ and $|\phi_0-\phi_b|\simeq 60^\circ$, consistent with the presence of strong magnetic-field rotations. In contrast, the non-switchback interval shows a much narrower angular distribution. The corresponding angular bounds are $|\phi_0-\phi_a|\simeq 20^\circ$ and $|\phi_0-\phi_b|\simeq 10^\circ$, indicating relatively small magnetic-field rotations around the mean field.

Among the 75 selected 8-hour intervals, 11 are classified as switchback intervals and 12 as non-switchback intervals. The remaining intervals exhibit maximum significant magnetic-field rotations between $25^\circ$ and $40^\circ$ and are therefore treated as an intermediate population. This intermediate population is analyzed separately to estimate its associated cascade rate and to provide a comparison with the switchback and non-switchback populations. In the next section, we describe the method used to estimate the turbulent cascade rate based on the PP98 exact law.

\subsection{Approach to estimate the cascade rate}\label{sec:approach}
We use the PP98 exact law to estimate the energy cascade rates associated with inward and outward Alfv\'enic turbulent fluctuations. PP98 derived an exact relation for the mixed vector third-order moment
\begin{equation}
\bf Y^\pm(\boldsymbol{\ell})=\left\langle 
\delta \mathbf{z}^{\mp}(\boldsymbol{\ell}) \, 
\left| \delta \mathbf{z}^{\pm}(\boldsymbol{\ell}) \right|^2 
\right\rangle 
\end{equation}
in the inertial range of homogeneous and incompressible MHD turbulence,
\begin{equation}
\nabla_{\boldsymbol{\ell}} \cdot 
{\bf Y}^\pm
= -4 \, \varepsilon^{\pm},
\end{equation} 

\noindent where $\varepsilon^{-}$ and $\varepsilon^{+}$ are the cascade rates associated with Alfv\'enic fluctuations propagating parallel and anti-parallel to the local background magnetic field $\bf B_\ell$, respectively. The Els\"asser increments are defined as $\delta \mathbf{z}^{\mp}=\mathbf{z}^{\mp}(\boldsymbol{\ell}+\boldsymbol{x})-\mathbf{z}^{\mp}(\boldsymbol{x})$, where $\boldsymbol{\ell}$ is the increment vector.

Applying Gauss's theorem to the above equation over a cylindrical volume, $\pi \ell_\parallel \ell_\perp^2$, in increment space gives
\begin{equation}
\int {\bf Y_\parallel^\pm}\cdot d{\bf S_\parallel}+\int {\bf Y_\perp^\pm}\cdot d{\bf S_\perp}
= -4\pi  \varepsilon^{\pm} \ell_\parallel \ell^2_\perp,
\end{equation}

\noindent where $d{\bf S_\parallel}$ and $d{\bf S_\perp}$ are the infinitesimal surface elements parallel and perpendicular to the local magnetic field, respectively, for a cylindrical volume of length $\ell_{\parallel}$ and radius $\ell_\perp$. Here we have $\boldsymbol{\ell}(\ell_\parallel,\ell_\perp,\theta)$ in cylindric coordinate.

The local mean magnetic field, $\bf B_\ell$, is defined as commonly used in previous studies,
\begin{equation}
{\bf B_\ell} = \frac{1}{2}({\bf B}(\bf {x+\boldsymbol{\ell}})+{\bf B}(\bf {x}))
\end{equation}

\noindent and is used to determine the local parallel and perpendicular directions.

In previous studies, the two terms on the left-hand side of the above surface-integral relation were commonly evaluated by assuming cylindrical symmetry of the cascade rate \citep{macbride05,macbride08,stawarz09,boldyrev09,stawarz10,stawarz11}. In the present analysis, we instead consider only pairs of points, $(\bf x,\bf x+\boldsymbol \ell)$, for which $\delta \mathbf{z}^{\mp}_\parallel=0$. This condition is appropriate for shear Alfv\'enic fluctuations, for which the fluctuations are transverse to the local magnetic field. Under this condition, ${\bf Y_\parallel^\pm}=0$, and the parallel contribution to the surface integral vanishes. The remaining perpendicular contribution becomes
\begin{equation}
\int {\bf Y_\perp^\pm}(\ell^\prime_\parallel,\ell_\perp,\theta)\cdot {\bf n} ~\ell_\perp~d\theta ~d{\ell^\prime_\parallel}
= -4\pi  \varepsilon^{\pm} \ell_\parallel \ell^2_\perp,
\end{equation}

\noindent where $d{\bf S_\perp}={\bf n} ~\ell_\perp~d\theta ~d{\ell^\prime_\parallel}$, and $\bf n$ is the unit vector along $\boldsymbol{\ell}$. This relation motivates the definition of the reduced, averaged mixed third-order moment ${\hat{ Y}_\perp^\pm}(\ell_\perp)$, which satisfies
\begin{equation}
{\hat{ Y}_\perp^\pm}(\ell_\perp)= -2 \varepsilon^{\pm} \ell_\perp,
\label{eq:main}
\end{equation}
where
\begin{equation}
{\hat{ Y}_\perp^\pm}(\ell_\perp)= \frac{1}{2\pi \ell_\parallel} \int {\bf Y_\perp^\pm}(\ell^\prime_\parallel,\ell_\perp,\theta)\cdot {\bf n} ~d\theta ~d{\ell^\prime_\parallel}.
\end{equation}
Here, ${\hat{ Y}_\perp^\pm}(\ell_\perp)$ is the reduced mixed third-order structure function. Empirically, it is obtained by averaging over the azimuthal angle $\theta$ and the parallel separation $\ell^\prime_\parallel$ using a large number of realizations:
\begin{equation}
{\hat{ Y}_\perp^\pm}(\ell_\perp)=\langle {\bf Y_\perp^\pm}(\ell^\prime_\parallel,\ell_\perp,\theta) \cdot {\bf n} \rangle_{\theta,\ell^\prime_\parallel}.\label{eq:Yreduced}
\end{equation}

\subsection{Empirical estimation of the cascade rate}

At this step, we involve the frozen-in-flow Taylor's approximation~\citep{taylor38} that connects the space lag vector $\boldsymbol{\ell}=-V_{0x} \tau~ \boldsymbol{\hat{x}}$ to the time lag $\tau$ through the dominant $x$-component of the mean velocity $\mathbf V_0$ calculated in each selected 8-hour-long intervals.

To determine the empirical values the cascade rates $\varepsilon^{\pm}$ in imbalanced Alfv\'enic turbulence in the slow solar wind for the selected intervals from \emph{Wind} we use equation~\eqref{eq:Yreduced} to evaluate the scaling of  ${\hat{ Y}_\perp^\pm}(\ell_\perp)$, which requires the evaluation of average ${\bf Y_\perp^\pm}(\boldsymbol{\ell})\cdot {\bf n}$ over $\ell_\parallel$ and $\theta$, with the consideration that ${\bf Y_\parallel^\pm}(\boldsymbol{\ell})\simeq0$. For that, we need to follow these steps:

{\it Step~1}: for each pair of points $t$ and $t+\tau$ in a selected interval, we evaluate the vector $\mathbf{\hat{ Y}^\pm}(t,\tau)=
\delta \mathbf{z}^{\mp}({t,\tau}) \, 
\left| \delta \mathbf{z}^{\pm}(t,{\tau}) \right|^2 
$, where $\delta \mathbf{z^\pm}({t,\tau})=\mathbf{z}^\pm(t+\tau)-\mathbf{z}^\pm(t)$. Also, we determine the local mean field $\mathbf{B}_l(t,\tau)=[\mathbf{B}_l(t+\tau)+\mathbf B_l(t)]/2$. This will allow us to estimate the local ratio {\bf $r(t,\tau)=\left|\delta z^\pm_\parallel(t,\tau)/\delta z^\pm_\perp(t,\tau)\right|$}, where $\parallel$ ($\perp$) refers to parallel (perpendicular) direction with respect to $\mathbf B_l(t,\tau)$. For that pair of points, we will be able to determine $\boldsymbol{\ell}(t,\tau)=-V_{0x} \tau~ \boldsymbol{\hat{x}}$ and its perpendicular projection $\boldsymbol\ell_\perp(t,\tau)=\boldsymbol{\ell}(t,\tau)-\ell_\parallel(t,\tau)~\boldmath e_B$, where $\mathbf e_B=\mathbf B_l(t,\tau)/B_l(t,\tau)$ is unit vector along $\mathbf 
B_l(t,\tau)$, and $\ell_\parallel(t,\tau)=\boldsymbol{\ell}(t,\tau)\cdot \mathbf e_B$ is the parallel component of $\boldsymbol{\ell}(t,\tau)$.

{\it Step~2}: At this point, we use equation~\eqref{eq:Yreduced} to estimate  ${\hat{ Y}_{\alpha,\perp}^\pm}(\ell_\perp)$ as a function of $\ell_\perp$ for each selected interval labeled with $\alpha$. First, we need to introduce $\delta \ell_\perp$ bin size for $\ell_{\perp,i}$ that is varying from  $\ell_\perp=|-V_{0x} \tau_{min}|$ to $|-V_{0x} \tau_{max}|$, where $\tau_{min}=5$ min and $\tau_{max}=1$ hour. Then, we perform a conditioned average over time for each bin $\ell_{\perp,i}$ and $\ell_{\perp,i}+\delta \ell_\perp$ as
\begin{equation}
{\hat{ Y}_{\alpha,\perp}^\pm}(\ell_{\perp,i})=\langle {\bf Y_\perp^\pm}(t,\tau) \cdot {\bf n}(t,\tau) \rangle^{cd}_{t},\label{eq:Yestimate}
\end{equation}

\noindent such that, only terms for the conditions {\bf $r(t,\tau)\le 0.2$} and $\ell_{\perp,i}\le\ell_\perp(t,\tau)<\ell_{i,\perp}+\delta \ell_\perp$ are considered in the sum in equation~\eqref{eq:Yestimate}. Choosing a relatively small ratio $r(t,\tau)$ will make the contribution of ${\bf Y_\parallel^\pm}(t,\tau)$ negligible which is consistent with the methodology described in section~\ref{sec:approach}.  A small value of the ratio, $r(t,\tau)<0.2$, is consistent with the predominantly Alfv\'enic nature of solar-wind turbulence and provides a sufficiently large number of pairs (more than two million) for a statistically robust estimation of the averaged quantities $\hat{Y}_\perp^\pm(\ell_\perp)$ over the selected intervals defined below. In contrast, imposing a more restrictive criterion, $r(t,\tau)<0.1$, substantially reduces the number of available pairs to approximately several hundred thousand, resulting in noisier estimates of $\hat{Y}_\perp^\pm(\ell_\perp)$.

{\it Step~3}: Once ${\hat{ Y}_{\alpha,\perp}^\pm}(\ell_\perp)$ is computed, we now estimate the reduced mixed third-order function ${\hat{ Y}_{SB,\perp}^\pm}(\ell_\perp)$ and ${{ Y}_{NSB,\perp}^\pm}(\ell_\perp)$ for switchback and non-switchback intervals, respectively, as
\begin{align}
\hat{Y}_{SB,\perp}^\pm(\ell_\perp) 
&= \left\langle \hat{Y}_{\alpha,\perp}^\pm(\ell_\perp) \right\rangle^{SB}_\alpha \\
\hat{Y}_{NSB,\perp}^\pm(\ell_\perp) 
&= \left\langle \hat{Y}_{\alpha,\perp}^\pm(\ell_\perp) \right\rangle^{NSB}_\alpha,
\end{align}
where $\langle ...\rangle^{SB}_\alpha$ ($\langle... \rangle^{NSB}_\alpha$) represents average over all selected intervals that with (without) field switchbacks.

{\it Step~4}: At this stage we can perform linear fits to $\hat{Y}_{SB,\perp}^\pm(\ell_\perp)$ ($\hat{Y}_{NSB,\perp}^\pm(\ell_\perp)$) with respect to $\ell_\perp$ as 
\begin{align}
\hat{Y}_{SB,\perp}^\pm(\ell_\perp)&=-2\epsilon^{\pm}_{SB} \ell_\perp \\
\hat{Y}_{NSB,\perp}^\pm(\ell_\perp)&=-2\epsilon^{\pm}_{NSB} \ell_\perp.
\label{eq:SB}
\end{align}

Then, the total heating rate $\epsilon^{\rm tot}_{SB}=(\epsilon^{+}_{SB}+\epsilon^{-}_{SB})/2 $ ($\varepsilon^{\rm tot}_{NSB}=(\epsilon^{+}_{NSB}+\varepsilon^{-}_{NSB})/2 $) for imbalanced Alfv\'enic turbulence  in the slow solar wind with (without) field switchbacks.




\section{Results\label{sec:result}}
\subsection{Empirical cascade rates}
In this section we show the main findings of our analysis of cascade rate in intervals with and without SBs. Instead of using $\epsilon^+$ and $\epsilon^-$ associated with $\hat{Y}_{\perp}^\pm(\ell_\perp)$ for SB and NSB intervals, we use $\epsilon^{out}$ and $\epsilon^{in}$ associated with $\hat{Y}_{\perp}^{out}(\ell_\perp)$ and $\hat{Y}_{\perp}^{in}(\ell_\perp)$, respectively.
Here we define any quantity $Q^{out}=Q^{-}$ and $Q^{in}=Q^{+}$ if the local background $\mathbf B_l$ is pointing outward from the sun and $Q^{out}=Q^{+}$ and $Q^{in}=Q^{-}$ if it is pointing inward.

\begin{figure}[!t]
    \centering
    \includegraphics[width=0.47\textwidth]{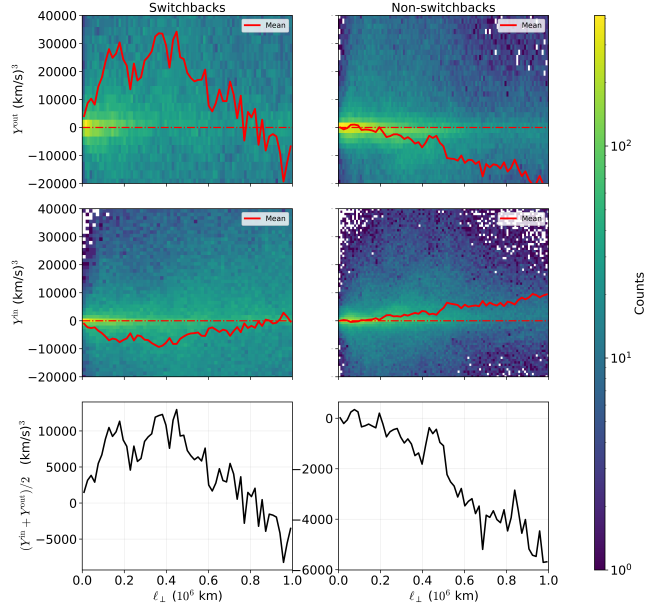}
     
    \caption{The top and middle rows show the distributions of the local estimates of the mixed third-order structure functions, $Y^{\rm out}(\ell_\perp)$ and $Y^{\rm in}(\ell_\perp)$, respectively, as a function of the perpendicular lag $\ell_\perp$. The color scale indicates the number of samples contributing to each bin, while the red curves denote the corresponding mean values. The bottom row displays the average quantity $Y=(Y^{\rm out}+Y^{\rm in})/2$. The quantities plotted here are calculated using the two intervals shown in Figure 1.}
    \label{fig:fig2}
\end{figure}

The quantities plotted in Figure 2 are calculated using the switchback and non-switchback intervals shown in Figure 1. Each interval may be regarded as a single realization of the turbulent fluctuations; therefore, it is not necessarily expected to accurately reproduce the asymptotic scaling of the mixed third-order structure functions, $Y_{\perp}^{\mathrm{out/in}}(\ell_\perp)$, shown in Figure 2. Because these estimates are obtained from individual intervals rather than from an ensemble average, we denote them by $Y^{\mathrm{out/in}}_\perp(\ell_\perp)$ throughout the paper, reserving the notation $\hat{Y}_{\perp}^{\mathrm{out/in}}(\ell_\perp)$ for the statistically converged quantities discussed later. 

The colored density maps in Figure~\ref{fig:fig2} show the distribution of the single-point measurements contributing to the estimate of $Y^{\mathrm{out/in}}_\perp(\ell_\perp)$, while the solid curves represent their corresponding mean values. The significant scatter observed at all scales highlights the large statistical variability associated with a single interval and demonstrates that the resulting scale dependence can deviate substantially from the expected linear behavior. 
\begin{figure}[!t]
    \centering
    \includegraphics[width=0.5\textwidth]{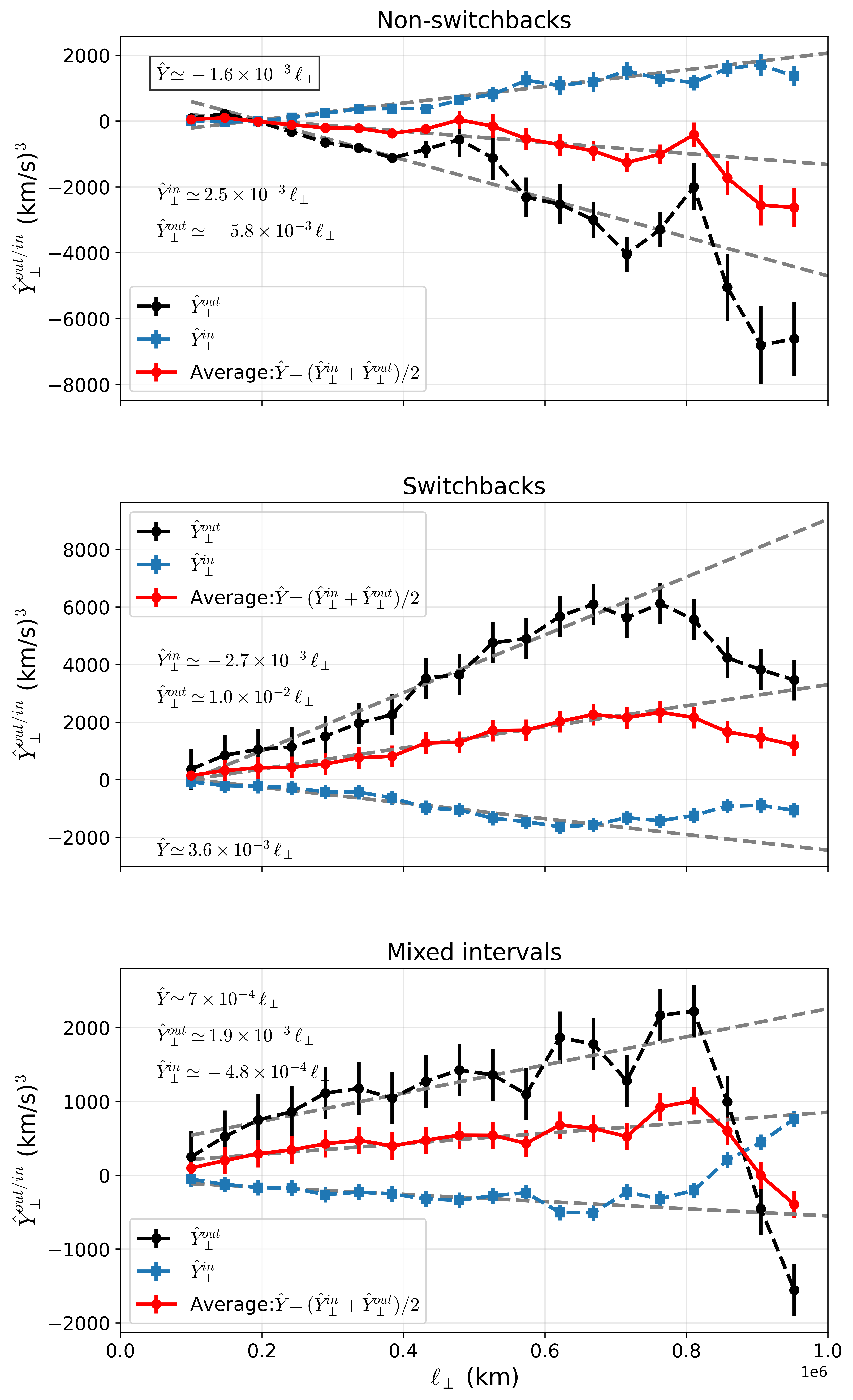}
     
    \caption{Ensemble-averaged structure functions associated with outward ($\hat{Y}_{\perp}^{\rm out}$; black dashed curves with circles) and inward ($\hat{Y}_{\perp}^{\rm in}$; blue dashed curves with squares) Alfv\'enic fluctuations as functions of the perpendicular lag, $\ell_\perp$. The red solid curves show the average quantity $\hat{Y}=(\hat{Y}_{\perp}^{\rm out}+\hat{Y}_{\perp}^{\rm in})/2$. The top, middle, and bottom panels correspond to non-switchback intervals, switchback intervals, and the combined (mixed) sample, respectively. Gray dashed lines indicate the linear fit scaling within the inertial range.}
    \label{fig:fig3}
\end{figure}  

Figure~\ref{fig:fig3} displays the ensemble-averaged third-order structure functions $\hat{Y}_{NSB,\perp}^{\rm out/in}(\ell_\perp)$ (top panel), $\hat{Y}_{SB,\perp}^{\rm out/in}(\ell_\perp)$ (middle panel), and $\hat{Y}_{mix,\perp}^{\rm out/in}(\ell_\perp)$ (bottom panel). The latter quantity is obtained by averaging over all selected intervals, including both switchbacks and non-switchbacks intervals, and therefore represents the mixed sample.

In addition to the inward and outward contributions, we consider the averaged mixed third-order structure function $\hat{Y}_\perp(\ell_\perp)=\left[\hat{Y}_{\perp}^{\rm out}(\ell_\perp)+\hat{Y}_{\perp}^{\rm in}(\ell_\perp)\right]/2$. As shown in Figure~\ref{fig:fig3}, $\hat{Y}_\perp(\ell_\perp)$ exhibits an approximately linear dependence on $\ell_\perp$ over a broad range of scales, consistent with the prediction of third-order turbulence theory. Linear fits to these curves yield $\hat{Y}_\perp \simeq
\left(-1.6\times10^{-3}\,\mathrm{km^2\,s^{-3}}\right)\ell_\perp$ for non-switchbacks intervals, $\hat{Y}_\perp\simeq \left(3.6\times10^{-3}\,\mathrm{km^2\,s^{-3}}\right)\ell_\perp$  for switchbacks intervals, and $\hat{Y}_\perp \simeq \left(0.7\times10^{-3}\,\mathrm{km^2\,s^{-3}}\right)\ell_\perp$ for the mixed sample. Here $\ell_\perp$ is given in km unit.

We further justify in Section~3.2 that the linear fits provide a good description of the curves in Figure~3 over a substantial portion of the inertial-range. The deviations from the linear fits occur primarily at $\ell_\perp \gtrsim 7\times10^{5}$~km, near the outer scale.

Using Equation (\ref{eq:main}) and using the linear scaling obtained from fits to the mixed third-order structure functions, we determine the empirical value of the total cascade rates $\epsilon^{\rm tot}$ for switchbacks, non-switchbacks and mixed intervals. The resulting values are  summarized in Table~\ref{tab:cascade_rates} For non-switchbacks intervals, characterized by relatively small magnetic-field rotations, the inferred cascade rate from the linear fit associated with outward-propagating fluctuations is positive, indicating a forward transfer of energy toward smaller scales. In contrast, the inferred cascade rate associated with inward-propagating fluctuations is negative, suggesting that the inward-propagating fluctuations contribute oppositely to the nonlinear energy transfer. Nevertheless, the total cascade rate remains positive with value $\epsilon^{\rm tot}_{\rm NSB}\simeq 0.8 \times 10^{3}$ J kg$^{-1}$ s$^{-1}$, implying that the net transfer of turbulent energy is directed from large to small scales, as expected in standard MHD turbulence.

A markedly different behavior is observed in switchbacks intervals. In this case, the inferred cascade rate of inward-propagating fluctuations is positive but comparatively weak, whereas the inferred cascade rate of outward-propagating fluctuations is negative and dominates the total contribution. As a result, the total cascade rate is negative with value $\epsilon^{\rm tot}_{SB}\simeq -1.8\times10^{3}$ J kg$^{-1}$ s$^{-1}$. Within the framework of third-order turbulence theory, this result is consistent with a reversal of the net energy transfer direction and therefore suggests the presence of an inverse transfer process associated with intervals containing large magnetic-field rotations. Such behavior may indicate that the turbulent dynamics within switchbacks are influenced by physical processes such as local instabilities, strong shear, magnetic reconnection, or other nonlinear interactions capable of modifying the conventional forward cascade.

The mixed sample exhibits a negative total cascade rate of about $-0.35\times10^3$ J kg$^{-1}$ s$^{-1}$, reflecting the combined contributions of switchbacks and non-switchbacks intervals and other intermediate intervals. The fact that the switchbacks and non-switchbacks populations contribute with opposite signs may help explain why previous studies have reported substantially different values, and in some cases negative values, of the solar-wind turbulent cascade rate. These results therefore suggest that the presence and relative abundance of switchbacks can have a significant impact on the measured energy-transfer properties of solar-wind turbulence.

\vspace*{1.0cm}
\hspace*{-1.4cm}
\setlength{\tabcolsep}{4pt}
\begin{table}[ht]
\centering
\begin{tabular}{l|ccc}
\hline
Intervals & $\epsilon^{\rm tot}$ & $\epsilon^{\rm out}$ & $\epsilon^{\rm in}$ \\
 & \multicolumn{3}{c}{ ($10^{3}$ J kg$^{-1}$ s$^{-1}$)} \\
\hline
NSB   & $(0.80\pm0.10) $ &
        $(2.90\pm0.30) $ &
        $(-1.20\pm0.10) $ \\
SB    & $(-1.80\pm0.10) $ &
        $(-5.0\pm0.20) $ &
        $(1.30\pm0.10) $ \\
Mixed & $(-0.35\pm0.05) $ &
        $(-0.95\pm0.16) $ &
        $(0.24\pm0.06) $ \\
\hline
\end{tabular}
\caption{The average cascade rate and the cascade rates associated with the inward- and outward-propagating Alfv\'enic fluctuations for the switchback and non-switchback intervals.}
\label{tab:cascade_rates}
\end{table}

\subsection{Power spectra of inward and outward Els\"asser fluctuations}

\begin{figure}[!t]
    \centering
    \includegraphics[width=0.5\textwidth]{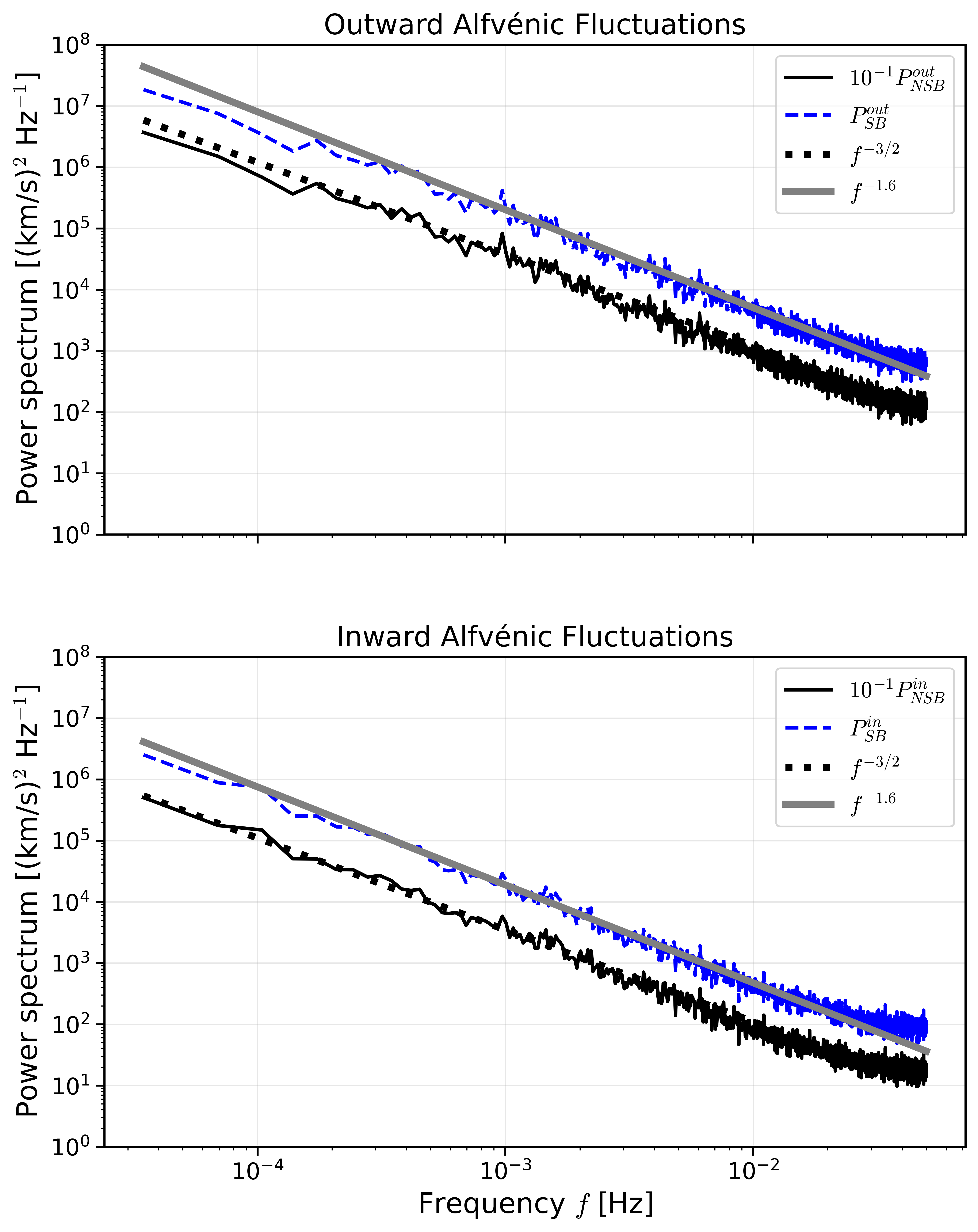}
     
    \caption{Power spectral densities of the outward, $P^{out}$ (top panel) and inward, $P^{in}$, (bottom panel) corresponding to Els\"asser fluctuations, $z^{\rm out}$ and $z^{\rm in}$, for switchback (SB; blue dashed curves) and non-switchback (NSB; black solid curves) intervals. The spectra are plotted as a function of frequency in the spacecraft frame. The dotted and solid reference lines indicate power-law scalings proportional to $f^{-3/2}$ and $f^{-1.6}$, respectively. The power spectra corresponding to the non-switchback intervals are multiplied by a factor of $0.1$ for better visibility. }
    \label{fig:fig4}
\end{figure}

To further characterize the turbulent properties of switchback and non-switchback intervals, we computed the ensemble-averaged power spectral densities of the outward- and inward-propagating Alfv\'enic fluctuations, represented by the Els\"asser variables ${\bf z}^{\rm out}$ and ${\bf z}^{\rm in}$, respectively. The resulting spectra are shown in Figure~\ref{fig:fig4} for both switchback (SB) and non-switchback (NSB) intervals.

For comparison, Figure~\ref{fig:fig4} also shows reference power-law scalings proportional to $f^{-3/2}$ and $f^{-1.6}$. In non-switchback (NSB) intervals, the power spectra of both ${\bf z}^{\rm out}$ and ${\bf z}^{\rm in}$ are generally more consistent with an $f^{-3/2}$ scaling than with the Kolmogorov-like $f^{-5/3}$ scaling. This result is in agreement with previous observational studies showing that imbalanced Alfv\'enic turbulence in the solar-wind tends to exhibit spectral indices closer to $-3/2$ \citep{chen20,bourouaine20}, consistent with the predictions of scale-dependent alignment theories of Alfv\'enic turbulence \citep{boldyrev06a}.

In contrast, in switchback intervals, where large magnetic-field rotations coexist with imbalanced Alfv\'enic fluctuations, the spectra are found to be slightly steeper, with spectral indices closer to $-1.6$. This value lies between the $-3/2$ scaling predicted by scale-dependent alignment models and the classical Kolmogorov $-5/3$ scaling. The observed steepening suggests that the presence of large magnetic-field rotations may modify the nonlinear turbulent cascade, leading to a systematic change in the spectral scaling. These results therefore indicate that the coexistence of large magnetic-field rotations in imbalanced Alfv\'enic turbulence can influence the spectral properties of the turbulent fluctuations as well as the cascade rates.

\noindent
The observed rollover of the third-order structure functions at large scales, 
$\ell_\perp \gtrsim 7\times10^{5}$~km in Figure~3, may be related to the transition 
toward the outer scale of the turbulence. In particular, the power spectra of the 
dominant outward-propagating Alfv\'enic fluctuations appear to deviate slightly from 
the $-5/3$ or $-3/2$ power-law scaling at frequencies below approximately 
$f_0\sim4\times10^{-4}$~Hz. This frequency is comparable to the break frequency 
associated with the outer scale reported by \cite{dorseth24a,dorseth24b} for Alfv\'enic 
turbulence in the slow solar wind near 1~au. Taylor's hypothesis has been shown 
to provide a reasonable connection between temporal and spatial scales near 1~au 
\citep[see, e.g.,][]{bourouaine20a}. Under this approximation, the corresponding 
perpendicular scale can be estimated as $L_\perp=V_0\sin\phi_0/f_0$, where $V_0$ 
is the solar-wind speed and $\phi_0$ is the angle between the mean magnetic field 
and the radial flow direction. Using a typical slow-solar-wind speed of 
$V_0=400$~km~s$^{-1}$, representative of the mean speed of our intervals, and a 
Parker-spiral angle of $\phi_0\simeq45^\circ$ near 1~au, we obtain 
$L_\perp\sim7\times10^{5}$~km. This scale is close to the perpendicular scale 
at which the rollover of the third-order structure functions begins, suggesting 
that the observed departure from linear scaling may be associated with the 
transition toward the outer scale of the turbulence.


\section{Conclusion and Discussion}
\label{sec:discussion}

We have presented an observational analysis of the turbulent cascade rate in imbalanced Alfv\'enic turbulence in the slow solar-wind using data from \emph{Wind}. The main objective was to determine whether the presence of magnetic-field switchbacks modifies the direction and magnitude of the turbulent energy transfer. To address this question, we selected intervals that are weakly compressible, approximately homogeneous, have low velocity shear, and exhibit high normalized cross-helicity. These intervals were then separated into switchback and non-switchback populations according to the angular spread of the magnetic-field direction around the background field.

The cascade rate was estimated using a reduced form of the Politano--Pouquet exact law. In contrast with previous approaches that often assume isotropic or axisymmetric turbulence, our method focuses on the perpendicular contribution to the mixed third-order structure function. This is achieved by conditioning the statistics on increments dominated by perpendicular Els\"asser fluctuations relative to the local magnetic field. This procedure is consistent with the shear-Alfv\'enic nature of the selected intervals and avoids imposing a global symmetry assumption on the turbulence.

Our results show that non-switchback intervals have a positive total cascade rate. This indicates a forward cascade of turbulent energy from large scales toward smaller scales, as expected for homogeneous MHD turbulence. In these intervals, the outward-propagating fluctuations provide the dominant positive contribution to the total cascade rate, while the inward-propagating component may show an inverse cascade due to the negative cascade rate. Nevertheless, the net energy transfer remains forward, suggesting that non-switchback Alfv\'enic turbulence (in the slow solar wind) behaves consistently with the classical turbulent cascade scenario.

In contrast, switchback intervals exhibit a negative total cascade rate. This negative value indicates an inverse transfer of energy from smaller scales toward larger scales. The inverse cascade is mainly associated with the outward-propagating Alfv\'enic fluctuations, whose contribution becomes negative and dominant in the presence of large magnetic-field rotations. This result suggests that switchbacks can substantially alter the energy-transfer process in Alfv\'enic turbulence in the slow solar wind.

The physical interpretation of this result is that switchbacks may not simply represent ordinary Alfv\'enic fluctuations embedded in the turbulent cascade. Instead, they may be associated with additional physical mechanisms that modify the scale-to-scale energy transfer. Possible mechanisms include magnetic-field folding, local shear-driven instabilities, reconnection-related structures, or nonlinear effects associated with the expansion and evolution of the solar wind. These processes may inject energy at intermediate or small scales, redistribute energy across scales, or locally reverse the direction of the cascade.

Our finding of a negative cascade rate in switchback intervals near 1 au appears to differ from the results reported by \cite{hernandez21}, based on Parker Solar Probe observations at heliocentric distances $R$ closer to the Sun ($R \lesssim 50,R_{\odot}$). Hernández et al. (2021) reported an enhanced forward (positive) cascade rate in switchback intervals, in contrast to our finding. However, the two studies employ different approaches to estimate the cascade rate. First, Hernández et al. assumed isotropic symmetry in the evaluation of the Politano--Pouquet relation, whereas our analysis accounts for the anisotropic nature of the turbulent structures. Second, they estimated the cascade rate separately for individual intervals, each spanning several hours, whereas in our analysis we estimated the cascade rate by averaging over all selected switchback intervals. In addition, our analysis is restricted to slow-wind intervals characterized by homogeneous, incompressible, and Alfv\'enic turbulence. We therefore speculate that differences in the analysis methodology and/or in the heliocentric regions investigated by the two studies may contribute to the discrepancy between the results.

The negative cascade rate in switchback intervals found in our analysis may also help explain previous observational studies that reported negative or weak cascade rates in highly Alfv\'enic solar-wind intervals \citep[see, e.g., ][]{smith09,stawarz09,stawarz10,vasquez18}. If such intervals contain a significant population of switchbacks or large magnetic-field rotations, the measured cascade rate may reflect not only the standard turbulent cascade but also the influence of these additional structures. Therefore, separating switchback and non-switchback intervals is essential for interpreting cascade-rate measurements in the solar wind.

Our analysis of the power spectra of the inward- and outward-propagating Els\"asser fluctuations further demonstrates that non-switchback intervals exhibit spectral indices close to the $f^{-3/2}$ scaling commonly observed in imbalanced Alfv\'enic turbulence, and switchback intervals display systematically steeper spectra, with spectral indices approaching $f^{-1.6}$. This result suggests that the large magnetic-field rotations characteristic of switchbacks can actively influence the cascade dynamics and the transfer of energy across scales.

Overall, our analysis supports the conclusion that the turbulent cascade in Alfv\'enic slow solar wind is significantly modified by large magnetic-field rotations. Intervals without large field rotations exhibit a forward cascade consistent with the standard picture of MHD turbulence, whereas intervals containing switchbacks display an inverse cascade. These findings suggest that switchbacks, or large magnetic-field rotations, are not simply a byproduct of standard homogeneous turbulence, but may instead arise from plasma instabilities or from the radial evolution of large-scale inhomogeneous structures.

\begin{acknowledgments}
SB is supported by NASA grants 80NSSC23K0776, and 80NSSC24K0137.
\end{acknowledgments}

\bibliographystyle{aasjournalv7}
\bibliography{MyLibrary.bib}

\end{document}